\documentclass[twocolumn,showpacs,preprintnumbers,amsmath,amssymb,superscriptaddress,10pt,aps,prb]{revtex4-1}
\usepackage{epsfig,amsopn}
\usepackage{graphicx}
\usepackage{epstopdf}
\usepackage{sidecap}
\usepackage{amsmath,amssymb}
\usepackage{amsthm}
\usepackage{enumerate}

\begin{document}
\title{Chiral nodes  and oscillations in the Josephson current in Weyl semimetals}
\author{Udit Khanna}
\affiliation{Harish-Chandra Research Institute, Chhatnag Road, Jhunsi, Allahabad 211 019, India.}
\author{Dibya Kanti Mukherjee}
\affiliation{Harish-Chandra Research Institute, Chhatnag Road, Jhunsi, Allahabad 211 019, India.}
\author{Arijit Kundu}
\affiliation{Physics Department, Technion, 320003, Haifa, Israel}
\affiliation{Department of Physics, Indiana University, Bloomington, IN 47405}
\author{Sumathi Rao}
\affiliation{Harish-Chandra Research Institute, Chhatnag Road, Jhunsi, Allahabad 211 019, India.}

\begin{abstract}
The separation of the Weyl nodes in a broken time-reversal symmetric Weyl semimetal  leads to helical quasi-particle excitations at the Weyl nodes, which, when coupled with overall spin conservation allows only inter-nodal transport at the junction of the Weyl semimetal with a superconductor. This leads to an unusual periodic oscillation in the Josephson current as a function of $k_0L$, where  $L$ is  the length of the Weyl semimetal and $2k_0$ is the inter-nodal distance. This oscillation is robust and should be experimentally measurable,  providing a direct path to 
confirming the existence of chiral nodes in the Weyl semimetal.
\end{abstract} 

\maketitle
\emph{Introduction.}---%
Weyl semimetals (WSM), which have received much interest recently due to their non-trivial transport characteristics, are  3D topological systems where conduction and valence bands touch at two or more `Weyl' points~\cite{Vishwanath2011,Burkov2011a,Burkov2011b,Zyuzin2012a,Hosur2012}. 
According to a no-go theorem~\cite{Nielsen1981}, gapless Weyl nodes in a WSM appear as pairs in momentum space with each of the nodes having a definite `chirality', a quantum number that depends on the Berry flux enclosed by a closed surface around the node. Gauss law prevents the annihilation of the nodes unless two of them with opposite chirality are brought together, which provides the `topological' protection of the Weyl nodes~\cite{Turner2013}. A WSM phase requires broken time-reversal and/or inversion symmetry and a growing number of systems has been put forward which realize the WSM phase~\cite{Xu2015,Lv2015,Liu2015}. 

\begin{figure}[!ht]
\centering
\includegraphics[width=0.45\textwidth]{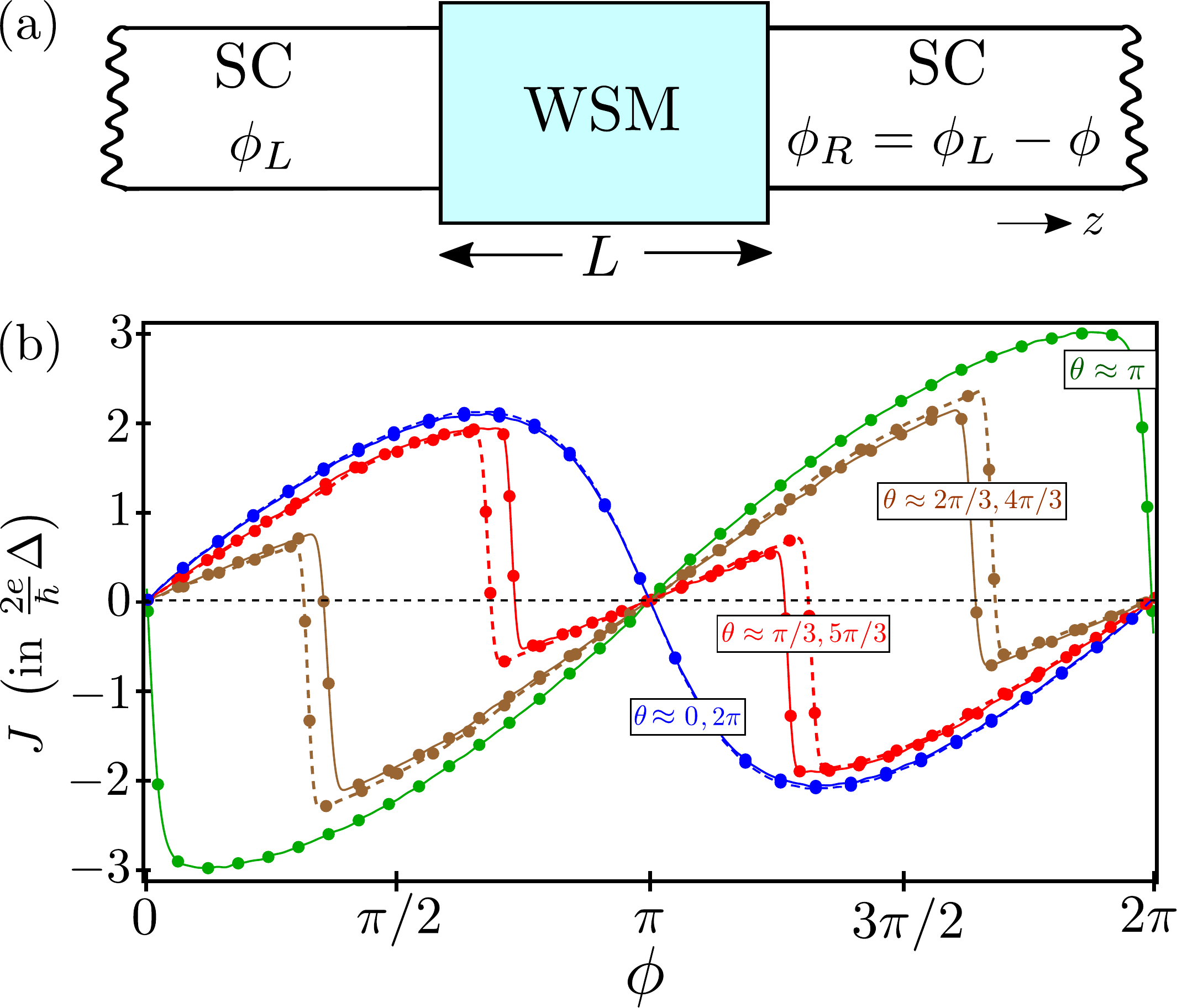}
\caption{ (Color online) (a) The setup for Josephson current with two superconductors (SC) characterized by phases $\phi_{\text{L}}$ and $\phi_{\text{R}}$ sandwiching a  WSM of length $L$ between them where a time-reversal broken perturbation separates the Weyl nodes in momentum space by $2k_0$ in $k_z$. (b) The Josephson current (at normal incidence) is periodic in $L$ with a period of $\pi/k_0$. We show the zero-temperature Josephson current, Eq.~(\ref{eq:JC}), as a function of the superconducting phase difference $\phi$, for various values of $L$ in solid (dashed) lines for $\theta$ between 0 and $\pi$ ($\pi$ and $2\pi$), where $\theta=2k_0L~\text{mod}(2\pi)$. The parameters used are $k_0L=31.4, 32.0,32.5, 33, 33.5, 34.0,34.454$, $\hbar^2k_0^2/2m_W=10\mu_W=10^3\Delta=\mu_S/2, ~p\ll k_0$ and $m_S=m_W$.}
\label{fig:fig1}
\end{figure}

The separation of the chiral nodes, allows charge pumping between the nodes in the presence of parallel electric and magnetic fields,
as a consequence of the chiral anomaly~\cite{Adler1969}, and this has led to detailed studies of transport in Weyl semi-metals in 
several recent  papers\cite{Vazifeh2013,Vazifeh2014,Hosurqi2013,Chen2013,Biswas2013,Ominato2014,Sbierski2014,Khanna2014, Burkov2015a,Son2013,Burkov2014,Burkov2015b,Gorbar2014,Goswami2015,
Ghimire2015,Klier2015,Zyuzin2012b,Chernodub2014,Cortijo2015,Burkov2015c,Chen2013,Baum2015}.

In this paper we study  the current in a simple Josephson junction setup, depicted in Fig.~\ref{fig:fig1}(a). The helical quasi-particle excitations at the Weyl nodes, due to the overall spin conserving processes at a WSM-superconductor (SC) junction, allow only inter-nodal transport~\cite{Uchida2014}.  Further, we show, unlike in a normal metal-SC interface, the inter-nodal `normal' (electron to electron) reflection process in a WSM-SC interface is not suppressed  even for energies close to Fermi-energy, due to the broken time-reversal symmetry separating the Weyl nodes. The Josephson current, flowing through the bound levels formed by multiple inter-nodal `normal' and Andreev (electron to hole) processes in a SC-WSM-SC system, consequently, acquires a specific periodicity as a function of the length of the WSM which depends only on the separation of the Weyl nodes in the momentum space (see Fig.~\ref{fig:fig1}(b)). We argue that both of these features are robust  because they are not only  bulk effects, but they are also protected by the robustness of the Weyl nodes. We also discuss the feasibility of experimental observations of this transition in our system, which can confirm the presence of chiral nodes in WSM.

This oscillation in the Josephson current and the resulting changes of sign of the critical current at arbitrary values of $\phi$ (or the $0$-$\pi$ transition) is an inherent property of the SC-ferromagnet-SC junction~\cite{Bulaevski1977,Buzdin1982,Buzdin1991,Rmpsfs} and has also been experimentally observed\cite{Ryazanov2001}. Since our model also explicitly violates time-reversal invariance, our results show quite a strong  similarity with the Josephson current in similar systems\cite{Golubov} as well in semiconductor nanowires with Zeeman coupling\cite{Yokoyama2014}. 

\emph{Model and geometry.}---%
We consider the geometry as shown in Fig.~\ref{fig:fig1}(a) with the superconductors at $z<0$ and $z>L$ and the Weyl semimetal (WSM) in the region $0<z<L$. We model the WSM starting from the standard Hamiltonian describing a 3D TI in the Bi$_2$Se$_3$ family~\cite{Fu2010,Qi2011}, regularized on a simple cubic lattice and adding a time-reversal breaking perturbation $b_z$ to access the WSM phase~\cite{Khanna2014} -
\begin{align}\label{eq:ham}
  H_{0} =& \epsilon_k \tau_x - \lambda_z \sin k_z \tau_y  \nonumber \\
  & -\lambda \tau_z \left( \sigma_x\sin k_y  - \sigma_y \sin k_x\right) + b_z \sigma_z.
\end{align}
Here $\epsilon_k = \epsilon - 2t \sum_i \cos k_i$ is the kinetic energy,  $\tau$ ($\sigma$) represent the orbital (spin) degrees of freedom and  $\lambda$, $\lambda_z$ are  the strengths of the spin-orbit coupling. 
In the limit $\lambda_z\ll M\ll b_z$, (where $M$ is defined as $\epsilon-6t$\cite{supple}), this simplifies to a two-band model for a WSM,  where,  in the absence of the spin-orbit coupling, $\lambda$, the bands have opposite spins~\cite{Uchida2014,supple}. The chiral fermion excitations around the two Weyl nodes (which we choose to be at $ (0,0,\pm k_0)$, where $t k_0^2 = b_z - M$) are described by the Hamiltonian
\begin{align}\label{eq:HWSM}
H_{\text{WSM}} = \epsilon_k\sigma_z -\mu_W + \lambda (k_x\sigma_x + k_y\sigma_y)
\end{align}
with $\epsilon_k = (\hbar^2/2m_W) (k_x^2+k_y^2+k_z^2-k_0^2)$ being the kinetic energy, $\mu_W$ being the chemical potential measured from the Weyl node and $m_W$ being the effective mass. Since the Weyl nature of the fermions is only evident at momenta which are small with respect to the symmetry breaking scale $k_0$, we choose the Fermi energy $\mu_W$ to be small enough, so that the Fermi surfaces around the two Weyl nodes are disconnected. The surface states for this model appear on the surfaces perpendicular to the $x$-axis and $y$-axis.  In this paper, we do not attempt to access transport through the surface states. Instead, we consider transport through the bulk of the WSM and hence restrict ourselves to transport parallel to the $z$-axis.

\begin{figure}[t]
\centering
\includegraphics[width=0.42\textwidth]{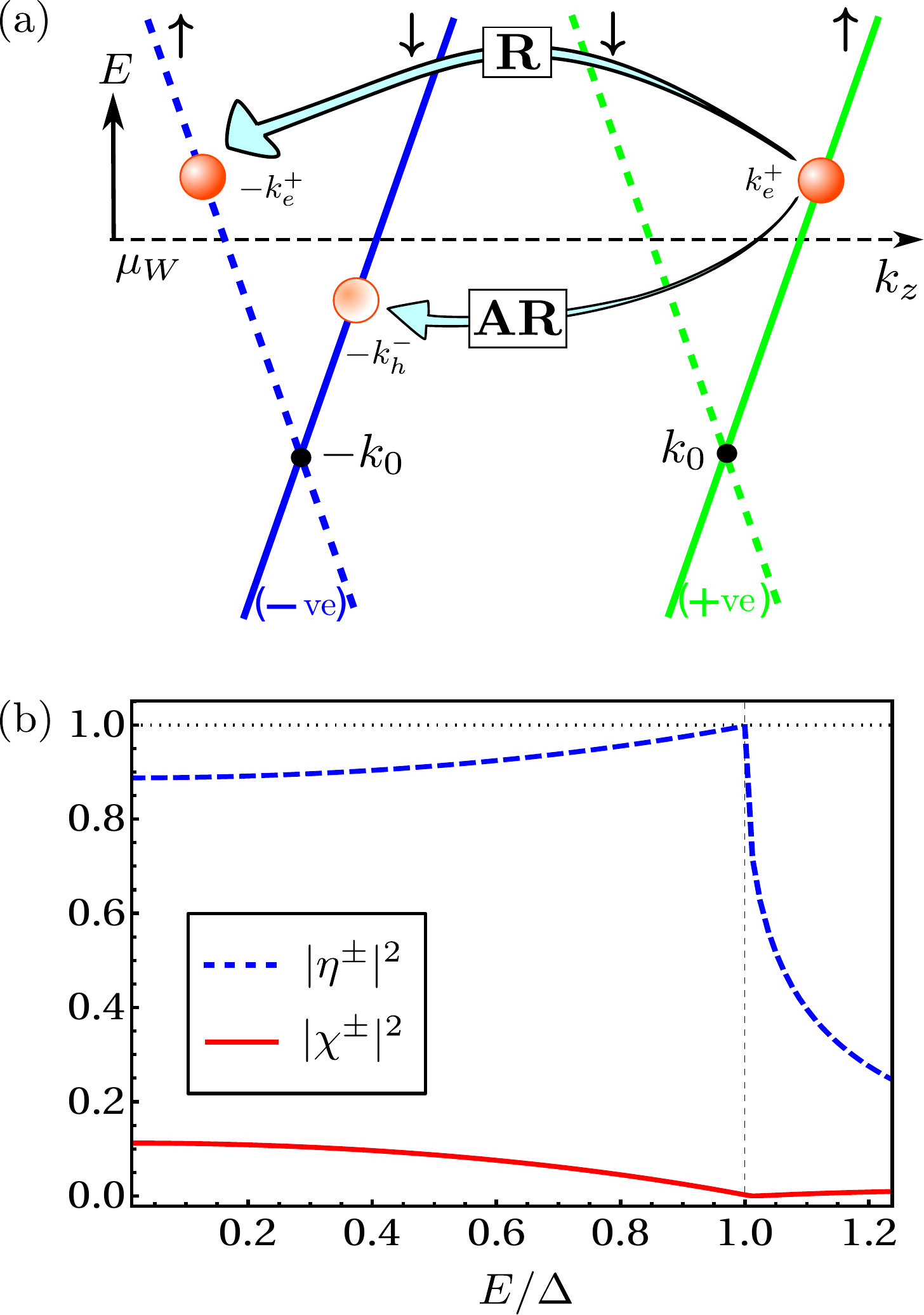}
\caption{ (Color online) (a) Both reflection (R) and Andreev reflection (AR) in WSM occur from one chiral node to another. The chiralities of the nodes are denoted as $+$ve and $-$ve, whereas the solid and the dashed lines show dispersions of Eq.~(\ref{eq:HWSM}) with positive and negative velocities ($= d E/dk$). The bands have opposite spins, which accounts for the change of chirality for both `normal' and Andreev reflection. (b) The probability that an electron will be reflected as an electron ($|\chi^{\pm}|^2$) or as a hole ($|\eta^{\pm}|^2$) at a WSM-SC interface, discussed in Eq.~(\ref{eq:simpleR}). Note that the probability of reflection as an electron is finite. The parameters used are the same as in Fig.~\ref{fig:fig1}(b).}\label{fig:fig2}
\end{figure}
The superconductors can be described in terms of the Boguliobov-de Gennes (BdG)
Hamiltonian  as 
\begin{align}\label{eq:SC}
 H_{\text{SC}}^j = \left(\begin{array}{cc} \xi_k I_{2\times 2}&  e^{i\phi_{j}}\Delta i  \sigma_y\\
-e^{-i\phi_j}\Delta i \sigma_y  & -\xi_k I_{2\times 2}\end{array} \right), 
\end{align}
where $\Delta$ is the pairing potential in the superconductor and $\xi_k = (\hbar^2 (k_x^2+k_y^2+k_z^2)/2m_S - \mu_S)$. $m_S$ is the effective mass of the electron in the superconductor and $\mu_S$ is the chemical potential. $\phi_j$ is the superconducting phase of the $j^{th}$ superconductor. For the
left and right superconductors, $j =L,R$. The parameter $\mu_S$ depends on the details of the superconducting material. In the numerical results shown, we consider $\mu_S \gg \Delta$, which is the  realistic limit. Also, for simplicity,  we consider $m_S \approx m_W$. 

\emph{WSM-SC junction.}---%
The solutions of Eq.~(\ref{eq:HWSM}) in the Nambu-Gor'kov space are now 4 component spinors. For incident energy $E$, the right-moving solutions with the wavefunctions proportional to $e^{\nu ik^{\nu}_{e}z}$ for electrons and $e^{-\nu ik^{\nu}_{h}z}$ for holes can be written in the basis of the two bands $\nu=\pm$, with   $$k_{e(h)}^{\nu} = \sqrt{k_0^2-p^2+\nu(2m_w/\hbar^2)\sqrt{(\mu_W+(-)E)^2-(\lambda p)^2}}$$ and  $p=\sqrt{k_x^2+k_y^2}$. The left-moving solutions can be written similarly with $k_{e(h)}^{\nu} \rightarrow -k_{e(h)}^{\nu}$. For the case of a WSM-SC junction, the WSM and and the superconducting wavefunctions on the two sides of the junction can be matched at the junction by  requiring the continuity of the wavefunction and its first derivative~\cite{supple}. This leads to the net reflection matrix $\mathcal{R}^j$ from the WSM-SC junction, which connects the left and right-moving solutions,
\begin{align}
\mathcal{R}^j = \left(\begin{array}{cc}
                    r_{ee}^j & r_{eh}^j \\
                    r_{he}^j & r_{hh}^j
                   \end{array}\right),
\end{align}
where the `normal' reflection matrices $r_{ee}^j(r_{hh}^{j})$, and the Andreev reflection matrices $r_{eh}^j(r_{he}^{j})$ denote, respectively, electron to electron (hole to hole) and hole to electron (electron to hole) processes at the interface with the $j$th superconductor.

For the purpose of physical interpretation, let us take the case of near-normal incidence ($k_0\gg p$) of an electron, where the reflection matrices reduce to the form:
\begin{align}\label{eq:simpleR}
 r_{ee}^j = \left(\begin{array}{cc}
                 \chi^+ & 0  \\
                 0 & \chi^{-} 
                \end{array}
 \right), \quad  r_{he}^j =e^{-i\phi_j} \left(\begin{array}{cc}
                 0 & \eta^+   \\
                 \eta^- & 0 
                \end{array}
 \right).
\end{align}
In this simplified form it is immediately clear that both the reflection and the Andreev reflection change the chirality (see  also Fig.~\ref{fig:fig2}(a)) and can only take place from one node to another  because of the chiral nature of the nodes. We plot the probabilities of normal and the Andreev reflection in Fig.~\ref{fig:fig2}(b). We note that even at energies close to the Fermi energy, normal reflection is not suppressed. The existence  of the new momentum scale $k_0 \neq k_F$, introduced by breaking the time-reversal symmetry, allows the  incident electron momentum  to be  different from the Fermi momentum  of the superconductor. This leads to the non vanishing of normal reflection~\cite{Sipr1997,supple}

\begin{figure}[t]
\centering
\includegraphics[width=0.46\textwidth]{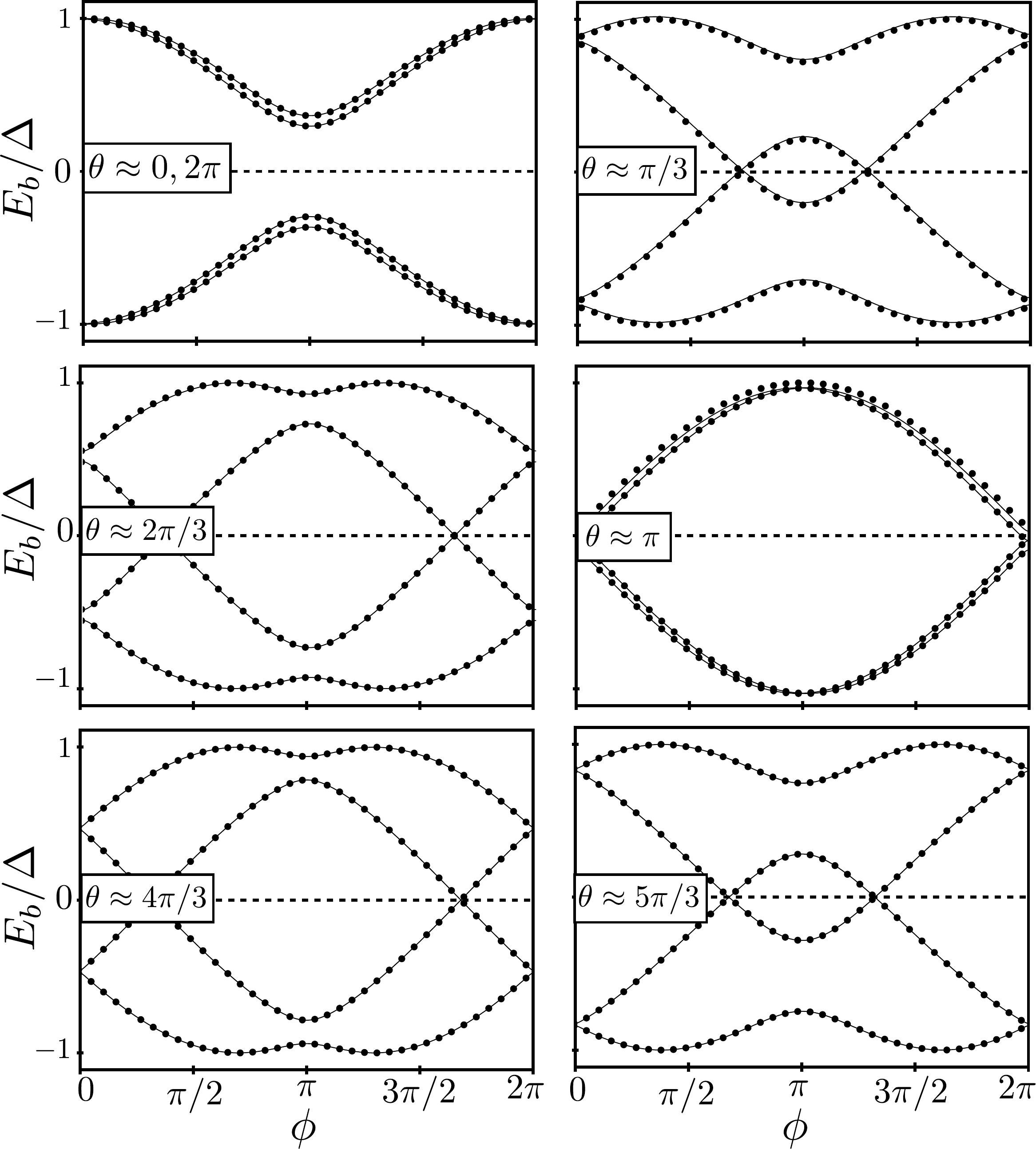}
\caption{The variation of the bound levels (solutions of Eq.~(\ref{eq:levels})) near the chemical potential with the length $L$ of the WSM for various values of $\theta$, where $ \theta=2k_0L ~\text{mod}(2\pi)$. The parameters used are the same as in Fig.~\ref{fig:fig1}(b).
}\label{fig:fig3}
\end{figure}

In contrast, note that for a topological insulator in 3D, the bulk is gapped and the non-trivial transport in junctions with superconductors is purely due to the surface states, where, the surface states consist of a Dirac metal with an odd number of nodes whose fermions have their spins aligned with the direction of motion (spin-momentum locking). This leads to  completely different physics for a topological insulator-superconductor junction~\cite{Fu2008, Soori2013}. Two dimensional graphene, on the other hand, is metallic and the transport is through the bulk. However, in graphene,  although there are two Dirac nodes (valleys), each of the nodes has fermions of both chiralities and the resulting process at the a superconducting interface is purely intra-nodal Andreev reflection~\cite{Beenakker2006,Kundu2010}.

\begin{figure}[t]
\centering
\includegraphics[width=0.45\textwidth]{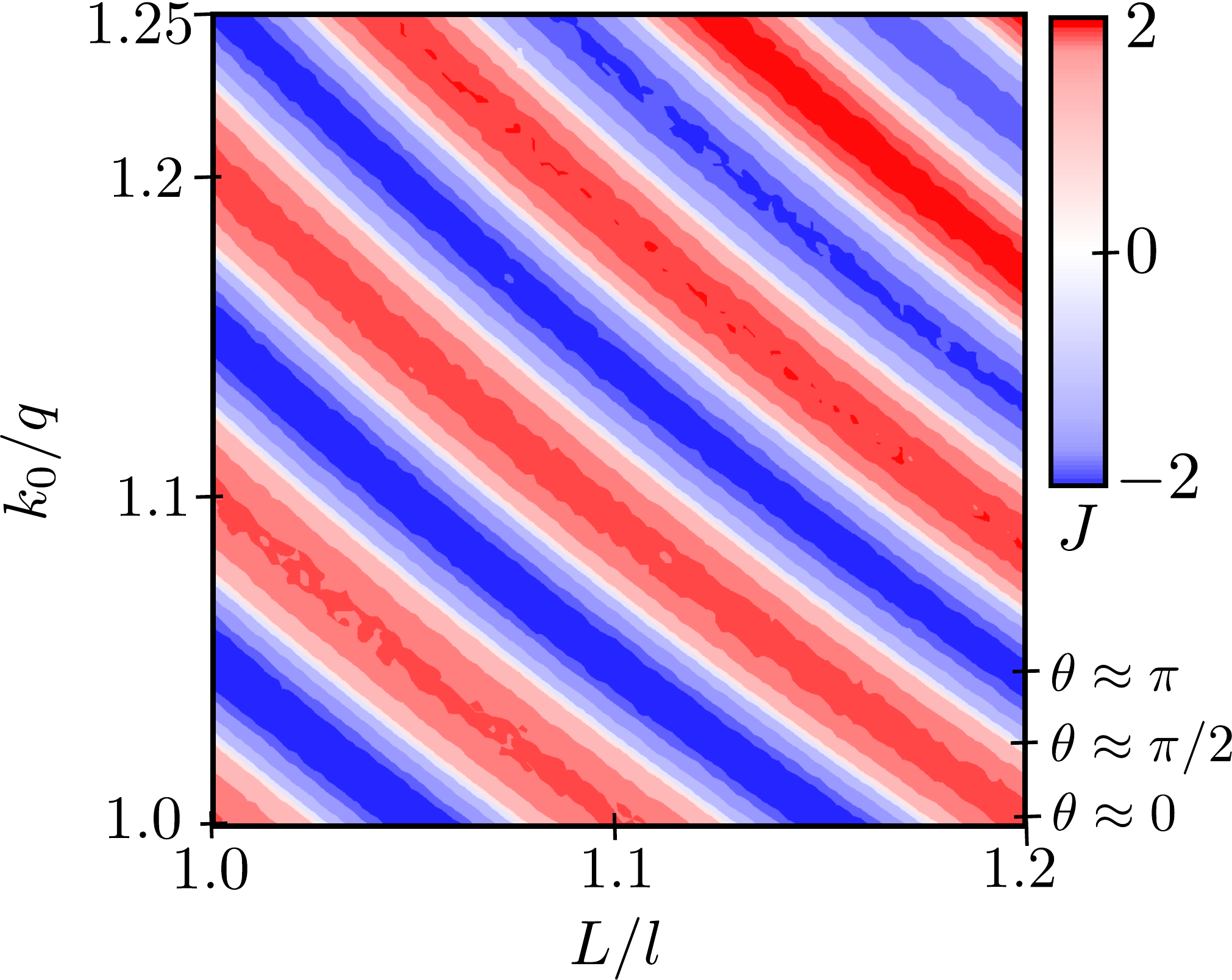}
\caption{(Color online) The Josephson current as a function of both $L$ and $k_0$ is shown at the value of $\phi\approx\pi/2$. The initial value at the origin is $(k_0,L)=( q,l),~ql\approx 10\pi$. The contours of constant current follow a set of (approximate) hyperbolas for constant $\theta=2k_0L~\text{mod}(2\pi)$, a few of which are shown in the right margin (with the minimum and the maximum current occuring near $\theta=\pi$ and $\theta=0$ respectively). Other parameters used are the same as in Fig.~\ref{fig:fig1}(b).}
\label{fig:fig4}
\end{figure}

\emph{Bound levels in the SC-WSM-SC geometry.}---%
Multiple reflections at the \emph{WSM-SC} boundaries lead to bound electronic levels in the \emph{SC-WSM-SC} geometry. But as discussed above, normal reflection amplitudes are not small at the \emph{WSM-SC} interface and, in general, there is no simple way of  summing up the amplitudes between the two superconductors to obtain the resonance condition when both Andreev and normal reflection amplitudes are non-zero. For the case of near normal incidence, however, the problem simplifies and the bound levels $E_b$ can be found  by solving
\begin{align}\label{eq:levels}
\det\left[I_{4\times4} - \mathcal{R}^L\mathcal{M}  \mathcal{R}^R\mathcal{M}\right]\left.\right|_{E=E_b}=0,
\end{align}
 where $\mathcal{M}$ is the matrix which accounts for the phase the electron/hole acquires while moving from one junction to another. We note that for $k_0\gg p$, Eq.~(\ref{eq:levels}) can still be used for approximate solutions. Writing
 \begin{align}
 \mathcal{R}^L\mathcal{M}  \mathcal{R}^R\mathcal{M} = \left(\begin{array}{cc}
                   \mathcal{T}_{ee} & \mathcal{T}_{eh} \\
                    \mathcal{T}_{he} & \mathcal{T}_{hh}
                   \end{array}\right),
\end{align}
in the limit of near normal incidence with $k_0^2/2m_W$ much larger than incident energy $E$ and $\mu_S$ much larger than pairing potential $\Delta$, the $\mathcal{T}$ matrices have the simplified form (with $m_S = m_W$):\cite{supple} 
  \begin{align}\label{eq:Tmat}
   & \quad \quad \mathcal{T}_{ee} = \left(\begin{array}{cc}
                 \alpha^+ & 0  \\
                 0 & \alpha^{-} 
                \end{array}
    \right), \quad   \mathcal{T}_{he} =\left(\begin{array}{cc}
                 0 & \beta^+   \\
                 \beta^- & 0 
                \end{array}
 \right) ,\\
 & \text{with}\quad \quad \alpha^{\pm} \approx e^{\pm 2ik_0L} \left(1+ 4iE\delta\right), \nonumber \\
&\quad \quad \quad \quad  \beta^{\pm} \approx \pm e^{\mp2ik_0L}2i(1+e^{-i\phi})\Delta ,\nonumber
  \end{align}
where $\delta=\sqrt{2m_W\mu_S}/k_0\Omega$ and $\phi=\phi_R-\phi_L$,. Also $\mathcal{T}_{hh}=\mathcal{T}_{ee}^*(E\rightarrow -E)$,  $\mathcal{T}_{eh}=\mathcal{T}_{he}^*(E\rightarrow -E)$. This immediately shows the periodicities of the $\mathcal{T}$ matrices,  $\mathcal{T}(\phi) = \mathcal{T}(\phi\rightarrow \phi + 2\pi)$ and $\mathcal{T}(2k_0L) = \mathcal{T}(2k_0L\rightarrow 2k_0L + 2\pi)$, which  implies that the bound levels $E_b$, the solutions of Eq.~\ref{eq:levels},  also inherit the same periodicities in $\phi$ and $2k_0L$. This additional periodicity of the levels with period $(\pi/k_0)$ in length  appears as a consequence of the inter-nodal normal and the Andreev reflections. The periodicities of $E_b$  in the difference of the superconducting phases $\phi$ and in $2k_0L$, in the limit of $k_0\gg p$ is shown in Fig.~\ref{fig:fig3}. This is our central result.

\emph{Periodic oscillations in the Josephson current.}---%
The Josephson current for  the system with the total Hamiltonian $H$ is written as $J_{\text{jos}}=\frac{2e}{\hbar}\left\langle \frac{\partial H}{\partial \phi}\right\rangle$, where the average is taken  over the states of the system. For the non-interacting system, where the length $L$ is much smaller than the coherence length in superconductors, the Josephson current flows through the bound levels (neglecting the continuum contribution) and can be estimated as~\cite{Chang1998}
\begin{align}\label{eq:JC}
 J(\mu_W)=\frac{2e}{\hbar}\sum_{b} \frac{\partial E_b}{\partial \phi}f(E_b-\mu_W),
\end{align}
where $f$ is the Fermi-distribution function. Apart from the $2\pi$ periodicity of the Josephson current in $\phi$, as the bound levels $E_b$ are periodic in $L$ with the periodicity of $\pi/k_0$, the Josephson current also inherits the same periodicity. This periodicity is shown explicitly in Fig.~\ref{fig:fig1}(b) for the case when $p=0$.

The periodic dependence in $L$ can also be written as an approximate periodicity in $k_0$ with a period of $\pi/L$. For large values of $L$, the rapid oscillations of $E_b$ with a small variation of $k_0$ outweighs any other dependence on $k_0$ and the periodicity is almost exact.  
The Josephson current as a function of both $k_0$ and $L$ is also shown in Fig.~\ref{fig:fig4}, where the locus of constant current approximately follows  $\theta=2k_0L~\text{mod}(2\pi)$. This is another of our main results.

\begin{figure}[t]
\centering
\includegraphics[width=0.45\textwidth]{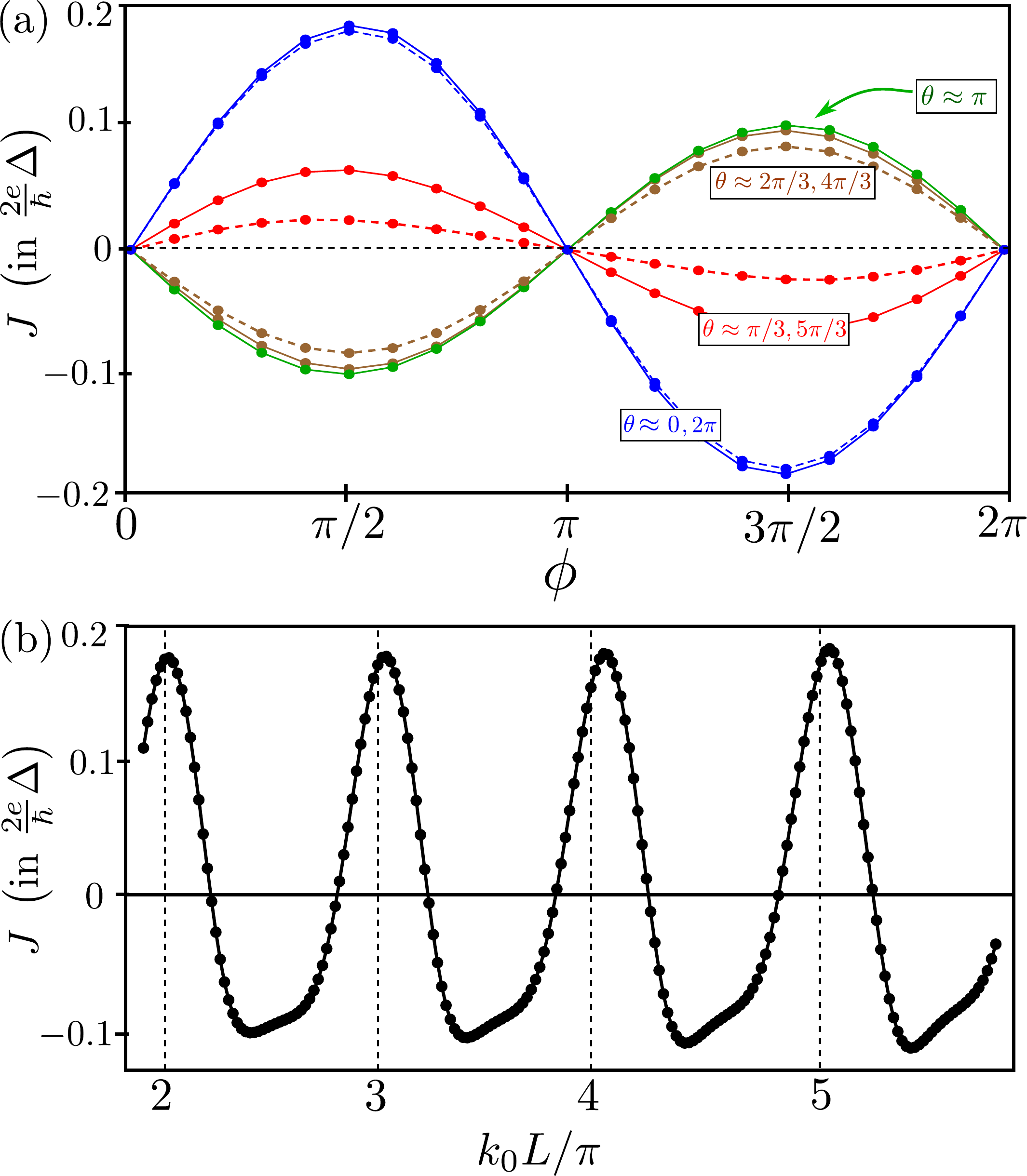}
\caption{(Color online) Result of lattice simulation: (a) Josephson current as a function of the superconducting phase difference $\phi$ for various values 
of $\theta=2k_0L ~\text{mod}(2\pi)$. (b) Josephson current as a function of the length  ($k_0 L/\pi$) at $\phi=\pi/2$. The parameters used 
are \cite{supple} $\tilde{t}=0.25t$ (see Eq.~(\ref{eq:coupling})), $\epsilon=6t$, $\lambda=\lambda_z=t$, $\Delta=0.01t$, $\mu=0.02t$ and $L=60$ sites. For these parameters, the position of Weyl nodes, $k_0 = b_z/\lambda_z$ (the lattice constant is taken as the unit of length).}
\label{fig:fig5}
\end{figure}

\emph{Lattice simulation.}---%
To analyse the case for non-normal incident angle $p\ne0$, we compute the Josephson current from the lattice version of Eq.~(\ref{eq:ham}) through its Green's function~\cite{supple,Vidal1994}.
The Green's function of the WSM, $g (\omega) = \left[(\omega + i \delta) \mathcal{I} - H_0\right]^{-1}$, is coupled with two superconductors $i=L,R$  
through the on-site self energy~\cite{Khanna2014}
\begin{align}\label{eq:coupling}
 \Sigma_i (\omega) = \frac{\tilde{t}}{\sqrt{\Delta^2 - \omega^2}} (\mathcal{I}_{\tau} + \tau^{x}) [\omega \mathcal{I}_{\zeta} - \Delta e^{i \phi_i} \zeta^{x}] \mathcal{I}_{\sigma}.
\end{align}
where $\zeta$ acts on the particle-hole degree of freedom in the Nambu basis and 
$\Sigma_i$ is defined only on the sites in contact with the $i^{\text{th}}$ superconductor. 
$\tilde{t}$ characterises the tunnelling between the superconductor and the WSM~\cite{Khanna2014}.
Then writing the full Green's function as $G (\omega) = (g^{-1}(\omega) - \Sigma_{L}(\omega) - \Sigma_{R}(\omega))^{-1}$ we compute the Josephson current~\cite{supple,Vidal1994}. We find that the  oscillations of current remain intact and this result has been summarized in Fig.~\ref{fig:fig5}. The bound levels $E_b$ can also be found approximately for $p\ll k_0$ using Eq.~(\ref{eq:levels}), and  the corresponding Josephson current also shows that the oscillation with $k_0L$ remains intact~\cite{supple}.


\emph{Feasibility of experimental realization.}---%
In the predicted WSM material $\text{Ta}\text{As}$~\cite{Huang2015b, Yang2015b}, chiral node pairs (formed by breaking inversion symmetry) are separated in momentum space by  a distance  $\sim0.02$\AA$^{-1}$. Assuming standard  electron mass, the relevant energy scale is about a milli electronvolt, which only becomes larger if the effective mass is smaller. Combining this with the fact that large momentum scattering (from $-k_0$ to $k_0$) is  needed to break the topological protection  of the chiral nodes, helical excitations in WSM are expected to be robust against disorder in a relatively clean sample. The periodicity of the Josephson current as well as the bound levels that we have discussed are, in principle, observable in tunneling experiments. The periodic variation of the bound-levels can also be probed in Andreev spectroscopy. For a typical sample, the length scales for such periodic variations would be of the order of few tens of nanometers. The separation of the Weyl nodes can also be tuned by adjusting the magnetic doping~\cite{Chen2010,Chang2013,Kurebayashi2014} to observe periodicities with the separation of the Weyl nodes.

The effect of having many Weyl nodes complicates the theoretical modeling and presents a weakness in our proposal. But, as long as the transport takes place along a pair of Weyl nodes, a similar periodicity in the Josephson current is expected.

\emph{Summary and conclusion.}---%
To summarize, we have shown explicitly, employing a simple model of the WSM, the occurrence of inter-nodal reflection processes at an WSM-SC interface due to spin conservation.
This gives rise to an unusual periodicity in the bound state spectra and consequently in the Josephson current that depends only on the separation of the two Weyl nodes and the size of the sample. This provides a direct path for possible observations of the manifestation of inter-nodal Andreev reflection in Weyl semimetals.

In closing, we sketch some problems for future studies. Apart from  transport signatures of the chiral anomaly in WSM, the appearance of surface states, and the consequent Fermi arc dispersion is another remarkable feature of the time-reversal broken WSM. Their transport characteristics in the Josephson current would be interesting to study. Finally, quantitative investigations of the effects of disorder and interactions on transport in the WSM are also  left for future studies.

A.~K. was supported in part by the NSF through Grant No.~DMR-1350663.


\newpage
\section*{Appendix}

\subsection{Model hamiltonian for the WSM}
In this section we derive Eq.~(2) near the Weyl nodes in Eq.~(1) of the main text. Starting with parent Hamiltonian Eq.~(1):
\begin{align}\label{eq:hams}
  H(k) =& \epsilon_k \tau_x - \lambda_z \sin k_z \tau_y  \nonumber \\
  & -\lambda \tau_z \left( \sigma_x\sin k_y  - \sigma_y \sin k_x\right) + b_z \sigma_z,
\end{align}
we define a mass term  $M = \epsilon - 6t$, so that, around the $\Gamma$ point 
\begin{equation} \epsilon_k = M + 2t \left( 3 - \sum_i \cos k_i\right) \approx M + t \sum_i k_i^2. \end{equation}
For $\lambda = 0 = b_z$ the hamiltonian (about $\Gamma$ point) is $ (M + t |k|^2) \tau_x - \lambda_z k_z \tau_y $, giving two doubly degenerate bands with the dispersion 
\begin{equation}
  \approx \pm \sqrt{M^2 + k^2_z (2Mt + \lambda_z^2) + 2 p^2 (Mt)},\nonumber
\end{equation} 
where $p^2 = k_x^2 + k_y^2$. Assuming that the mass is large enough, we can write 
\begin{equation}
  \pm \sqrt{M^2 + k_z^2 (2Mt + \lambda_z^2) + 2 p^2 (Mt)} \approx \pm (M + \alpha k_z^2 + t p^2),\nonumber
\end{equation}
where $\alpha = t + \lambda_z^2/2M$. 
This quadratic dispersion with a non-zero gap ($2M$) at the $\Gamma$ point is the starting point in Ref.~\onlinecite{Uchida2014}. Motivated by this, we define a unitary matrix $U$ such that
\begin{equation}
  U^{\dagger} \left[(M + t |k|^2) \tau_x - \lambda_z k_z \tau_y \right] U = (M + \alpha k_z^2 + t p^2) \tau_z. \nonumber
\end{equation}
In this new basis the full hamiltonian $U^{\dagger}H(k)U$ is
\begin{widetext}
\begin{align}
   \left( \begin{array}{cccc}
    M + b_z + \alpha k_z^2 + t p^2 & 0 & 0 & \lambda (k_y + i k_x) \\
    0 & M - b_z + \alpha k_z^2 + t p^2 & \lambda (k_y - i k_x) & 0 \\
    0 & \lambda (k_y + i k_x) & -M + b_z - \alpha k_z^2 - t p^2 & 0 \\
    \lambda (k_y - i k_x) & 0 & 0 & -M - b_z - \alpha k_z^2 - t p^2 
  \end{array} \right)~.
\end{align}
\end{widetext}
Clearly, $b_z$ lifts the spin degeneracy and shifts the bands up or down in energy while $\lambda$ mixes the uppermost and lowermost bands and the two bands in between. If the chemical potential is not large, the lowest energy excitations are only in the two bands close to zero energy. Limiting ourselves to the lowest energy excitations, we get the effective 2-band model
\begin{equation}
   \left( \begin{array}{cc}
    t (\alpha' k_z^2 + p^2 - k_o^2)_W & \lambda (k_y - i k_x) \\
    \lambda (k_y + i k_x) & -t (\alpha' k_z^2 + p^2 - k_o^2)
  \end{array} \right),
\end{equation}
where $t k_o^2 = b_z - M$ (assuming $b_z > M$) and $\alpha' = \alpha/t = 1 + \lambda_z^2/2Mt \approx 1$. This is the hamiltonian that we use in the main text.

\subsection{Solving the WSM-SC interface}
In this section,  we provide the details for the derivation of the reflection matrix in a WSM-SC system. Following Ref. 35 (Uchida et al.), the wavefunction of energy $E_i$ in the WSM is given by the following solutions of Eq.~(1) of the main text in the Nambu-Gor'kov space (with the Hamiltonian in the hole space written as $-H_{\text{WSM}}^*(-\mathbf{k})$):
\begin{align}
\psi_{\text{WSM}}(z) =\sum_{\sigma=\pm}\Bigg\{& \mathcal{E}^{\sigma} \left(a^{\sigma}_Re^{\sigma ik_e^{\sigma}z} +a^{\sigma}_Le^{-\sigma ik_e^{\sigma}z} \right) \nonumber \\
 +&\mathcal{H}^{\sigma} \left(b^{\sigma}_Re^{-\sigma ik_h^{\sigma}z} +b^{\sigma}_Le^{\sigma ik_h^{\sigma}z} \right)\Bigg\},
\end{align}
where $\sigma$ is the band index, $a(b)$ denotes the electron (hole) amplitude and $L(R)$ denotes the  left (right) moving solution.  $\mathcal{E}^{\sigma}(\mathcal{H}^{\sigma}$) are normalized eigenvectors, which are non-zero in electron (hole) sector of the Hamiltonian. In each sector $\mathcal{E}(\mathcal{H})^+\propto(f_{e(h)},(-)\lambda_{+(-)})^T$, and  $\mathcal{E}(\mathcal{H})^-\propto((-)\lambda_{-(+)},f_{e(h)})^T$, with $f_{e(h)} = \mu_W +(-)E_{i}+\sqrt{(\mu_W +(-)E_i)^2 - (\lambda p)^2}$, $\lambda_{\pm} = \lambda(k_x + i k_y)$. 

In the superconductor, the solutions of Eq.~(2) of the main text are: 
\begin{align}
\psi_{\text{SC}}(z)=\left(\begin{array}{c} uc_{\uparrow}\\ uc_{\downarrow}\\ -vc_{\downarrow}\\ vc_{\uparrow}\end{array}\right) e^{ i q_ez} +\left(\begin{array}{c} vd_{\downarrow}\\ -vd_{\uparrow}\\ ud_{\uparrow}\\ ud_{\downarrow}\end{array}\right) e^{ -i q_hz}, \nonumber
\end{align}
where, with $\Omega = \sqrt{\Delta^2-E_i^2}$,
$$ u(v) = \sqrt{\left(E_i +(-)i\Omega\right)/2E_i}$$
and $q_e$ and $-q_h$ are, respectively, the outgoing electron and hole momenta in the superconductor, defined as (with Fermi momentum $k_F$)
\begin{align}
 q_{e(h)} = \sqrt{k_F^2-p^2+(-)2m_S i \Omega/\hbar^2} ~.\nonumber
\end{align}

The boundary conditions at $z=0$ are given by the continuity of the wavefunction and its derivative:
\begin{align}
& \psi_{\text{WSM}}(0)=\psi_{\text{SC}}(0) \nonumber \\
& m_S\left(\begin{array}{cc} \sigma_z & 0\\ 0 & \sigma_z\end{array}\right)\partial_z\psi_{\text{WSM}}(z)\mid_{z=0} = m_W\partial_z\psi_{\text{SC}}(z)\mid_{z=0},\nonumber
\end{align}
with $\sigma_z$ being the Pauli matrix. By solving them one gets the reflection matrices,
\begin{align}\label{eq:Smat}
 &\left(\begin{array}{c}
        a_L^+\\
       a_L^-\\
        b_L^+\\
        b_L^-\\
       \end{array}
\right) = \left(\begin{array}{cc}
                 r_{ee} & r_{eh} \\
                 r_{he} & r_{hh} \\
                \end{array}
 \right) \left(\begin{array}{c}
        a_R^+\\
        a_R^-\\
        b_R^+\\
       b_R^-\\
       \end{array}
\right).
\end{align}
\subsection{Non-vanishing normal reflection}
For the simplest case,  if we take the only incident amplitude to be nonzero as $a_L^+=1$ on the left side of Eq.~(\ref{eq:Smat}), 
we can show that the amplitude of normal reflection is
\begin{align}
 a^+_R &= \frac{uc_{\uparrow}}{2}\left(1-\frac{q_e}{k_e^+}\right) + \frac{vd_{\downarrow}}{2}\left(1+\frac{q_h}{k_e^+}\right) \nonumber \\
 \text{with}~~ ud_{\downarrow} &= vc_{\uparrow}\left(1-\frac{q_e}{k_e^-}\right)/\left(-1-\frac{q_h}{k_e^-}\right).
\end{align}
In a normal metal (with normal incidence), $q_e\approx q_h\approx k_e^{\pm}\approx k_F$ when the incident energy is small, so that the amplitude of the normal reflection $a_R^+$ vanishes. However, in the WSM, $q_e\approx q_h\approx k_F$, but $k^{\pm}_e \approx k_0\neq k_F$. Hence, for the WSM, the normal reflection is not suppressed in general at small incident energies.

\vspace{0.2cm}
\subsection{Bound state spectrum}
The bound state spectrum for the SC-WSM-SC geometry can be found by using the reflection matrix at a single WSM-SC interface defined in Eq (S6). At near normal incidence, the bound states are given by the zeroes of the determinant
\begin{align}\label{eq:levelss}
\det\left[I_{4\times4} - \mathcal{R}^L\mathcal{M}  \mathcal{R}^R\mathcal{M}\right]\left.\right|_{E=E_b}=0, 
\end{align}
The $\mathcal{R}^L$ and $\mathcal{R}^R$ matrices are the reflection matrices at {\it SC-WSM} and {\it WSM-SC} interfaces respectively and the $\mathcal{M}$ matrix is the phase picked up by the electrons and holes in the WSM region (of length $L$). At near normal incidence, we can compute the matrices analytically and write
 \begin{align}
 \mathcal{R}^L\mathcal{M}  \mathcal{R}^R\mathcal{M} = \left(\begin{array}{cc}
                   \mathcal{T}_{ee} & \mathcal{T}_{eh} \\
                    \mathcal{T}_{he} & \mathcal{T}_{hh}
                   \end{array}\right),\nonumber
\end{align}
where ($\phi=\phi_R-\phi_L$),
\begin{align}
 &\mathcal{T}_{ee} = \left(\begin{array}{cc}
                \alpha_1^+ + \alpha_2^+ & 0  \\
                 0 & \alpha_1^{-} + \alpha_2^-
                \end{array}
 \right), \nonumber \\
&\mathcal{T}_{he} =\left(\begin{array}{cc}
                 0 & \beta_0^+(e^{-i\phi} \beta_1^+ +\beta_2^+)  \\
                \beta_0^-(e^{-i\phi} \beta_1^- + \beta_2^-) & 0 
                \end{array}
 \right).\nonumber
\end{align}
\begin{widetext}

\begin{align}
& \alpha_1^{\pm} = \frac{4e^{\pm i(k_e^\pm +k_h^\mp)L} e^{-i\phi} 
k_e^\pm k_h^\mp (q_e +q_h)^2 u^2 v^2}
{\left[ (k_e^\pm + q_e)(k_h^\mp +q_h) u^2 - 
(k_h^\mp - q_e)(k_e^\pm - q_h)v^2 \right]^2},
\nonumber \\
& \alpha_2^{\pm} = e^{\pm 2ik_{e/h}^\pm L}\frac{
\left[(k_e^\pm - q_e)(k_h^\mp + q_h)u^2 + (-k_h^\mp + q_e)(k_e^\pm + q_h) v^2\right]^2}
{\left[ (k_e^\pm + q_e)(k_h^\mp +q_h) u^2 - 
(k_h^\mp - q_e)(k_e^\pm - q_h)v^2 \right]^2},
\nonumber
\end{align}
\begin{align}
& \beta_0^{\pm} = \mp 2 k_e^{\mp} (q_e+q_h) uv, 
\nonumber \\
& \beta_1^{\pm} = e^{\mp i(k_e^\mp +k_h^\pm)L}\frac{
\left[(k_e^\mp + q_e)(k_h^\pm - q_h)u^2 - (k_h^\pm + q_e)(k_e^\mp - q_h) v^2\right]}
{\left[ (k_e^\mp + q_e)(k_h^\pm +q_h) u^2 - 
(k_h^\pm - q_e)(k_e^\mp - q_h)v^2 \right]^2},
\nonumber\\
& \beta_2^{\pm} = e^{\mp 2i k_e^\mp L}\frac{
\left[(k_e^\mp - q_e)(k_h^\pm + q_h)u^2 - (k_h^\pm - q_e)(k_e^\mp + q_h) v^2\right]}
{\left[ (k_e^\mp + q_e)(k_h^\pm +q_h) u^2 - 
(k_h^\pm - q_e)(k_e^\mp - q_h)v^2 \right]^2},
\end{align}
\end{widetext}
and $\mathcal{T}_{hh}=\mathcal{T}_{ee}^*(E\rightarrow -E)$,  $\mathcal{T}_{eh}=\mathcal{T}_{he}^*(E\rightarrow -E)$. This immediately shows the periodicities of the $\mathcal{T}$ matrices,  $\mathcal{T}(\phi) = \mathcal{T}(\phi\rightarrow \phi + 2\pi)$ and $\mathcal{T}(2k_0L) \approx \mathcal{T}(2k_0L\rightarrow 2k_0L + 2\pi)$ considering $E\ll k_0^2/m_W$, i.e, $k_{e(h)}^{\pm}\approx k_0$. The expressions can be simplified in the limit $\mu_S \gg \Delta$ and $k_0^2/2 m_W \gg E$ (the incident energy) and are given in the main text. 

\begin{figure}
\centering
\includegraphics[width=0.45\textwidth]{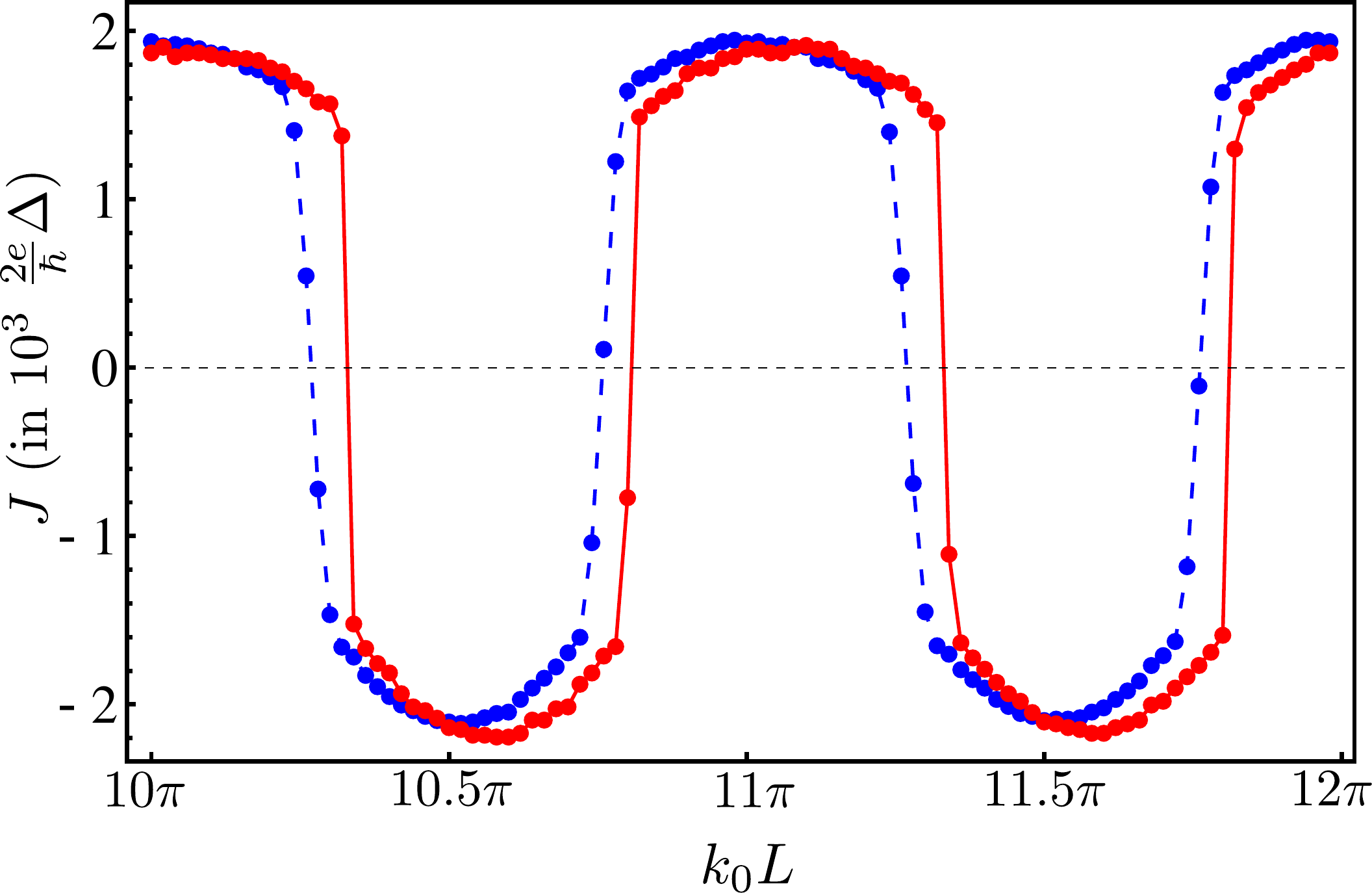}
\caption{(Color online) The zero-temperature  Josephson current at a fixed superconducting phase difference $\phi=\pi/2$ with the length of the WSM, $L$, keeping $k_0$ the same for transverse momenta $k_x,k_y=0$ in the blue (dashed) curve, while having the transverse momenta $k_x = 0.1, k_y=0$ in the red (solid) curve. Other parameters are the same as in Fig.~1(b) of the main text.}
\label{fig:S1}
\end{figure}

\subsection{Variation of Bound levels for $p\neq 0$}
We show a plot of the zero-temperature Josephson current with varying length of the WSM in Fig.~\ref{fig:S1} for transverse momentum  $p\ne0$ (but still one order of magnitude smaller than $k_0$). For the approximate determination of the bound levels $E_b$, we assume that Eq.~(6) of the main text still remains valid in this range of parameters. The figure shows that the periodicity with $k_0L$ is intact, which implies the robustness of the periodicity in non-normal but small-angle incidence of electron.

\subsection{Green's function formalism of Josephson current}
Following Ref.~\onlinecite{Vidal1994}, if $N_i$ represents the number operator of the $i$th site in the metallic system (WSM), then $-e\dot{N}_i$ can be written as the sum of the current flowing from the ($i-1$)th site to the $i$th site and from the $i$th site to the $(i+1)$th site, i.e, $\dot{N}_i = (-1/e)(J_{i-1\rightarrow i}+J_{i+1\rightarrow i})$. Each of these terms represents the Josephson current flowing in the system and is independent of the site $i$ for a large enough system. To elaborate, let us write the lattice Hamiltonian for the WSM in the following form (in the basis of spin operator $\sigma_z$ with eigenvalues $\sigma$ and parity operator $\tau_z$ with eigenvalues $\tau$):
\begin{align}
H = H^1+H^2& = \sum_{i,\sigma,\sigma',\tau,\tau'} a^{\dagger}_{i\sigma\tau}h^{(1)}_{i,\sigma\sigma'\tau,\tau'}a_{i\sigma'\tau'} \nonumber \\
 &+ \sum_{\langle ij\rangle,\sigma,\sigma',\tau,\tau'} a^{\dagger}_{i\sigma\tau}h^{(2)}_{ij,\sigma\sigma'\tau,\tau'}a_{j\sigma'\tau'}. \nonumber
\end{align}  
Then first term does not contribute to the current. In our case the second term of the Hamiltonian is (assuming translation invariance in the directions perpendicular to that of the flow of Josephson current)
\begin{align}
H^2 = a_{i}^{\dagger}(- t_{\text{WSM}}\tau_x + i\lambda_z\tau_y)\sigma_0a_{j}+~\text{h.c.}~.
\end{align}
And the current from the $i$th to the $(i+1)$th site as
\begin{align}
J_i(t) &= -\frac{ie}{\hbar}\sum_{\sigma,\tau} (t_{\text{WSM}}-\tau\lambda_z) \times \nonumber \\ &\Big( \langle a^{\dagger}_{i\sigma\tau}(t)a_{i+1\sigma\bar{\tau}}(t)\rangle - \langle a^{\dagger}_{i+1\sigma\bar{\tau}}(t)a_{i\sigma\tau}(t)\rangle \Big)
\end{align}
where $\bar{\tau} = -\tau$.
The above averages can be written in terms of the Green's function as
\begin{widetext}
\begin{align}
\mathbf{G}_{ij,\sigma\sigma',\tau\tau'}^{+-}(t,t') = i\left(\begin{array}{cc}
\langle a^{\dagger}_{i\sigma\tau}(t')a_{j\sigma'\tau'}(t)\rangle & \langle a_{i\sigma\tau}(t')a_{j\sigma'\tau'}(t)\rangle\\
\langle a^{\dagger}_{i\sigma\tau}(t')a^{\dagger}_{j\sigma'\tau'}(t)\rangle & \langle a_{i\sigma\tau}(t')a^{\dagger}_{j\sigma'\tau'}(t)\rangle\end{array}
\right)~.\nonumber
\end{align}
\end{widetext}
In the absence of any applied voltages the correlation function $\mathbf{G}^{+-}(\omega)$ is given by $\mathbf{G}^{+-}(\omega) = f(\omega) [\mathbf{G}_{\text{A}}(\omega)-\mathbf{G}_{\text{R}}(\omega)]$. $f(\omega)$ is the Fermi function and $\mathbf{G}_{\text{A}/\text{R}}$ are the advanced and retarded Green's functions.

\end{document}